\documentclass[onecolumn,pre,floats,aps,amsmath,amssymb,nofootinbib]{revtex4-2}
\usepackage{geometry}
\usepackage{graphicx}
\usepackage{epstopdf}

\newcommand{\be}{\begin{equation}}
\newcommand{\ee}{\end{equation}}
\newcommand{\bea}{\begin{eqnarray}}
\newcommand{\eea}{\end{eqnarray}}

\begin{document}

\title{Multiple dispersive bounds. I) The z-expansion}

\author{Silvano  Simula}  
\affiliation{Istituto Nazionale di Fisica Nucleare, Sezione di Roma Tre,\\ Via della Vasca Navale 84, I-00146 Rome, Italy}
\author{Ludovico  Vittorio}  
\affiliation{Physics Department and INFN Sezione di Roma, Universit\`a di Roma La Sapienza, P.le A.\,Moro 2, I-00185 Roma, Italy}

\begin{abstract}

We propose the implementation of two ingredients in the phenomenological applications of the unitary approach based on the $z$-expansion of hadronic form factors, commonly referred to as the Boyd-Grinstein-Lebed (BGL) $z$-expansion\,\cite{Boyd:1994tt, Boyd:1995cf, Boyd:1995sq, Boyd:1997kz}. The first ingredient is the explicit addition of a {\em unitarity filter} applied to a given set of input data for the hadronic form factors. This further constraint is not usually taken into account in the phenomenological applications of the BGL $z$-expansion. We show that it follows from the equivalence between the BGL approach and the Dispersion Matrix (DM) method\,\cite{DiCarlo:2021dzg}, which also describes hadronic form factors in a completely model-independent and non-perturbative way. The second ingredient is represented by the introduction of suitable kernel functions in the evaluation of unitarity bounds, leading to the application of {\em multiple dispersive bounds} to hadronic form factors, whenever data and/or (non-)perturbative techniques allow to do so. This idea may be useful for the investigation of many physical processes, from the analysis of the electromagnetic form factors of mesons and baryons to the study of weak semileptonic decays of hadrons. An explicit numerical application will be presented in the companion paper\,\cite{Simula:2025fft}, where  the effects of sub-threshold branch-cuts are analyzed. 

\end{abstract}

\maketitle

\section{Introduction}
\label{sec:intro}

Hadronic form factors are physical quantities which encode all the information about the strong non-perturbative dynamics among quarks and gluons. The knowledge of the precise behaviour of these functions is absolutely crucial in the investigation of many physical processes, from the analysis of electromagnetic form factors of mesons and baryons to the study of weak semileptonic decays of hadrons. An explicit example of current interest is offered by the study of semileptonic $B \to D^* \ell \nu_\ell$ decays, where a lot of results from Lattice QCD (LQCD) \cite{FermilabLattice:2021cdg, Harrison:2023dzh, Aoki:2023qpa} and from experiments \cite{Belle:2018ezy, Belle:2023bwv, Belle-II:2023okj, Belle:2023xgj} are now available. Exploiting the properties of unitarity and analyticity in this channel is, thus, of capital importance for phenomenological applications, in particular for the extraction of the Cabibbo-Kobayashi-Maskawa matrix element $\vert V_{cb} \vert$ from measurements and for the computation of the theoretical expectation value of the ratio of branching fractions $R(D^*)\equiv\Gamma(B \to D^* \tau \nu_{\tau})/\Gamma(B \to D^* \ell \nu_\ell)$ ratio (where $\ell=e,\,\mu$)  \cite{Martinelli:2023fwm, Ray:2023xjn, Kapoor:2024ufg, Bordone:2024weh, Duan:2024ayo, Li:2024weu, Martinelli:2024bov, Vittorio:2025txp, Bordone:2025jur}.

A parametrization widely used in the literature to study hadronic form factors in a way consistent with unitarity and analyticity is the $z$-expansion approach commonly referred to as the Boyd-Grinstein-Lebed (BGL) expansion\,\cite{Boyd:1994tt, Boyd:1995cf, Boyd:1995sq, Boyd:1997kz}. It can be applied to the description of the momentum dependence of hadronic form factors in their whole kinematical range and it is based on first-principles only. This parametrization allows to impose the properties of unitarity and analyticity on the form factors through the implementation of a suitable dispersive bound, which can be rephrased as a specific unitarity constraint on the coefficients of the BGL expansion itself.

In the present work and in the companion one\,\cite{Simula:2025fft}, we propose two main improvements of the unitary analysis, based on the BGL $z$-expansion, usually adopted in the literature for reproducing a given set of input data.
\begin{itemize}
\item We obtain a filter based on unitarity to be applied to the known input dataset for the form factors. This constraint is usually neglected in the literature, and it is conceptually different from the one that guarantees the unitarity of the BGL expansion. In this work we show how these two unitarity constraints can be distinguished one from each other and how they should be applied together.
\item We generalize the concept of the dispersive bound of Refs.\,\cite{Meiman63, Okubo:1971jf, Okubo:1971my, Okubo:1971wup} to the one of \emph{multiple} dispersive bounds through the introduction of suitable kernel functions in the evaluation of the so-called susceptibilities\,\cite{Boyd:1995sq, Bourrely:1980gp}, which can be estimated from the derivatives of appropriate two-point Green functions evaluated in momentum space. In this way, one can impose at the same time several dispersive constraints, whose effect on the final description of the form factors is to obtain more accurate predictions than the one resulting from the application of a single, total dispersive bound. This issue will be exploited from the numerical point of view in the companion paper\,\cite{Simula:2025fft}, in which multiple dispersive bounds will be used in the simple form of \emph{double} dispersive bounds, applied to the description of the effects of sub-threshold branch-cuts.
\end{itemize}

This first paper is organized as follows. In Section \ref{sec:BGL} we review the basic properties of the BGL $z$-expansion and, then, discuss the different constraints dictated by unitarity both on the coefficients of the BGL expansion and on the given input dataset for the form factors. To this end we derive again an additive representation of the BGL expansion, which turns out to be equivalent to the prediction of the DM approach. In Section \ref{sec:seminclusive}, we introduce the idea of multiple dispersive bounds and discuss how they can be implemented in the BGL expansion in absence of sub-threshold branch-cuts. The case in which sub-threshold branch-cuts are present is analyzed separately in the companion paper\,\cite{Simula:2025fft}. Our conclusions are summarized in Section \ref{sec:concl}, while Appendices\,\ref{sec:outer} and \ref{sec:additive} contain some details concerning aspects of the unitary BGL framework.

\section{The BGL expansion}
\label{sec:BGL}

In this Section we briefly recall some basic features of  the BGL $z$-expansion. For sake of simplicity, we consider the case of a single, generic form factor $f(t)$, where $t$ is the squared four-momentum transfer, describing the transition induced by a current $J$ (which may be the weak $V - A$ current or the electromagnetic vector current) between two hadrons $H_1$ and $H_2$ with masses $m_1$ and $m_2$, respectively.

Unitarity implies (see, e.g., Ref.\,\cite{Caprini:2019osi}) that the form factor $f(t)$ is an analytic function in the complex $t$-plane cut on the real axis from the lowest threshold $t_{th}$  to infinity\footnote{When (isolated) real poles due to bound states are present below the threshold $t_{th}$, a product of appropriate Blaschke factors can be introduced to guarantee analyticity (see later on).}. The discontinuity across the cut is twice the imaginary part of $f(t)$ along the cut and it is dictated by unitarity. The form factor $f(t)$ is real for real values of $t$ below $t_{th}$, fulfilling in this way the so-called Schwarz reflection property.
The above analytical properties imply that the form factor $f(t)$ may be represented through a dispersion relation of the form
\be
     \label{eq:DR}
     f(t) = \frac{1}{\pi} \int_{t_{th}}^\infty dt^\prime \frac{\mbox{Im}f(t^\prime)}{t^\prime - t - i \epsilon} ~ , ~
\ee
where the $i \epsilon$ defines the integral for values of $t$ on the branch-cut. Subtractions may be required to make the integral in Eq.\,(\ref{eq:DR}) convergent. We stress that in the present work we limit ourselves to the case in which the threshold $t_{th}$ is given by the pair-production threshold, namely
\be
     \label{eq:pair}
     t_{th} = t_+ \equiv (m_1 + m_2)^2 ~ , ~
\ee
where an on-shell pair $\overline{H}_1 H_2$ can be created from the vacuum by the current $J$. The case $t_{th} < t_+$, i.e.\,when sub-threshold branch cuts may be present, is discussed in the companion paper\,\cite{Simula:2025fft}.

The BGL expansion of Ref.\,\cite{Boyd:1997kz} is based on the introduction of the conformal mapping defined in terms of the variable
\be
    \label{eq:zplus}
    z = z(t; t_0) \equiv \frac{\sqrt{t_+ - t} - \sqrt{t_+ - t_0}}{\sqrt{t_+ - t} + \sqrt{t_+ - t_0}} ~, ~ 
\ee
where $t_0 < t_+$ is an auxiliary variable, which fixes simply the value of $t$ at which $z(t_0; t_0) = 0$. For real values of $t$ up to the threshold $t_+$ the conformal variable $z$ is real. In this regime, namely for $t \leq t_+$, the conformal variable ranges from $z=-1$ at $t = t_+$ to $z=1$ for $t \to - \infty$. On the branch-cut, namely for $t \geq t_+$, one has, instead, $|z| = 1$. We adopt the convention in which values of the conformal variable $z$ within the unit disk (i.e.\,$|z| \leq 1$) stay in the first Riemann sheet of the square root. A fundamental property is that $f(z)$, namely the form factor expressed in terms of the conformal variable $z$, eventually multiplied by the appropriate Blaschke factors (see below), is an analytic function of the real type in the unit disk, i.e. it is analytic and satisfies the Schwarz reflection principle $f(z^*) = f^*(z)$. The latter property implies that $f(z)$ is real for real values of $z$. 

The BGL expansion of the form factor $f(z)$ on the unit disk reads as\,\cite{Boyd:1997kz}
\be
    \label{eq:BGL}
    f(z) = \frac{\sqrt{\chi^U}}{\phi(z) B(z)} \sum_{k=0}^\infty a_k z^k ~ , ~ 
\ee
where $\phi(z)$ is a kinematical function, analytic of the real type and without zeros inside the unit disk, while $\chi^U$ is an {\em upper bound} to the quantity $\chi[f]$, associated to the form factor $f(z)$ as 
\be
    \label{eq:chi_plus}
    \chi[f] \equiv \frac{1}{2\pi i} \oint_{|z| = 1} \frac{dz}{z} |\phi(z) f(z)|^2 = \frac{1}{2\pi} \int_{-\pi}^\pi d\theta |\phi(e^{i\theta}) f(e^{i\theta})|^2 > 0 ~ , ~
\ee
namely
\be
    \label{eq:upper}
   \chi[f] \leq  \chi^U ~ . ~ \\[2mm]
\ee
Since the monomials $z^k$ are orthonormal when integrated on the unit circle, the inequality\,(\ref{eq:upper}) implies that the BGL coefficients $a_k$ appearing in Eq.\,(\ref{eq:BGL}) must satisfy the constraint
\be
    \label{eq:unitarity}
    \sum_{k=0}^\infty a_k^2 \leq 1 ~ . ~
\ee

In Eq.\,(\ref{eq:BGL}) the quantity $B(z)$ is a product of Blaschke factors related to (isolated) poles corresponding to bound states with masses $m_R$ below the threshold $\sqrt{t_+}$, namely
\be
    \label{eq:Blaschke}
    B(z) = \Pi_R \frac{z - z_R}{1 - z \, z_R} = \Pi_R \, z(t; m_R^2) ~ , ~
\ee
where $z_R = z(m_R^2; t_0)$ is real and $|z_R| < 1$.
The Blaschke product is an analytic function of the real type inside the unit disc, so that also the product $\phi(z) B(z) f(z)$ is analytic of the real type inside the unit disk. This implies that the coefficients $a_k$ in the expansion\,(\ref{eq:BGL}) are real. Furthermore, the Blaschke product is unimodular on the unit circle (i.e. $|B(z)| = 1$ for $|z| = 1$) and, therefore, it does not contribute to Eq.\,(\ref{eq:chi_plus}) at all.

\subsection{The dispersive susceptibility as the upper  bound $\chi^U$}
\label{sec:dispersive_bound}

A crucial point of the BGL expansion\,(\ref{eq:BGL}) is the estimate of the upper bound $\chi^U$ to $\chi[f]$, defined in Eq.\,(\ref{eq:chi_plus}). Such quantity depends both on the form factor $f(z)$ and on the function $\phi(z)$ on the unit circle, namely along the unitarity branch-cut. 
Thanks to the pioneering works of Refs.\,\cite{Meiman63, Okubo:1971jf, Okubo:1971my, Okubo:1971wup}, in many cases of physical interest it is possible to choose the function $\phi(z)$ so that the quantity $\chi^U$ can be estimated from derivatives of suitable two-point Green functions evaluated in momentum space (see Refs.\,\cite{Boyd:1995sq, Bourrely:1980gp}). In such cases, the explicit knowledge of the form factor $f(z)$ on the unit circle is not required for evaluating $\chi^U$.

To be more specific, the idea is to consider the polarization function $\Pi(q^2)$ associated to a double insertion of the hadronic current $J$. For instance, in the case of the weak vector or axial flavor-changing currents the polarization tensor $\Pi_{\mu \nu}(q^2)$ is defined as
\bea
    \label{eq:pol_tensor}
    \Pi_{\mu \nu}(q^2) & = & i \int d^4x \, e^{i q \cdot x} \, \langle 0 | {\cal{T}} \{ J_\mu^\dagger(x) J_\nu(0) \} 
                                           | 0 \rangle  ~ \nonumber \\[2mm]
                                 & = & q_\mu q_\nu \, \Pi_L(q^2) + \left(q_\mu q_\nu - g_{\mu \nu} q^2 \right) \,
                                           \Pi_T(q^2) ~ ,~
\eea
where ${\cal{T}}$ is the time-ordered product and $\Pi_{L(T)}(q^2)$ are the longitudinal(transverse) polarization functions.
The polarization function $\Pi(q^2)$, which may be $\Pi_L(q^2)$ or $\Pi_T(q^2)$ in Eq.\,(\ref{eq:pol_tensor}), satisfies a dispersion relation and its imaginary part on the branch-cut, dictated by unitarity, is given by a sum over a complete set of {\em on-shell} states $\overline{H}_1 H_2$ that couple the current $J$ to the vacuum. Using crossing symmetry and the positivity of each term in the completeness sum one arrives to a bound for the form factor $f(t)$ of the form\,\cite{Boyd:1997kz}
\be
     \label{eq:pair_bound}
     \int_{t_+}^\infty dt \, W_n(t) |f(t)|^2 \leq \chi_n ~ , ~ 
\ee
where $W_n(t)$ is a known weight function depending on the particular form factor $f(t)$. In Eq.\,(\ref{eq:pair_bound}) the quantity $\chi_n$ is a derivative of the polarization function, the so-called susceptibility, namely (cf., e.g., Ref.\,\cite{DiCarlo:2021dzg})
\be
     \label{eq:chi_n}
     \chi_n(q_0^2) = \left[  \frac{1}{n!} \frac{\partial^n}{\partial(q^2)^n} q^2 \Pi(q^2) \right]_{q^2 = q_0^2} = \frac{1}{\pi}\int_{t_+}^\infty dt \, \frac{t \, \mbox{Im}\Pi(t)}{(t - q_0^2)^{n+1}} ~ , ~
\ee
where $n$ is the order of the derivative connected to the number of subtractions required to make the dispersion integral finite and $q_0^2 < t_+$ is an auxiliary value of the squared four-momentum transfer. In this way, through the weight function $W_n(t)$ the absolute value of the kinematical function $\phi$ is completely specified along the branch-cut, i.e. on the unit circle. To extend it inside the unit disk one can interpret $\phi$ as an outer function or, more simply, one can follow the procedure of Ref.\,\cite{Bharucha:2010im}, based on the following unimodular substitutions
 \bea
       \label{eq:substitutions}
       \sqrt{t} & \to & \sqrt{\frac{-t}{z(t; 0)}} = \sqrt{t_+ - t} + \sqrt{t_+} ~ , ~ \\[2mm]
      \sqrt{t_- - t} & \to & \sqrt{\frac{t_- - t}{z(t; t_-)}} = \sqrt{t_+ - t} + \sqrt{t_+ - t_-} ~ , ~ \nonumber 
\eea
where $t_- \equiv (m_1 - m_2)^2$. 
Eqs.\,(\ref{eq:substitutions})  guarantee the analyticity of the function $\phi$ and the absence of zeros inside the unit disc.

In the case of the form factors relevant for the semileptonic decays $M_1 \to M_2 \ell \nu_\ell$, with $M_1$ being a pseudoscalar meson and $M_2$ a pseudoscalar or vector meson, the final expression for the kinematical function $\phi$ at $q_0^2 = 0$ reads as\,\cite{Boyd:1997kz, Bharucha:2010im}
\bea
    \label{eq:outer}
    \phi(z) & \to & \delta ~ \sqrt{\frac{n_I}{K \pi}} \left( \frac{t_+ - t}{t_+ - t_0} \right)^{1/4} \left( \sqrt{t_+ -t} + \sqrt{t_+ - t_0} \right) (t_+ - t)^{a/4}  \nonumber \\
               & \cdot & \left( \sqrt{t_+ -t} + \sqrt{t_+ - t_-} \right)^{b/2} \left( \sqrt{t_+ -t} + \sqrt{t_+} \right)^{-c-3} ~ , ~
\eea
where $n_I$ is an isospin Clebsch-Gordan factor, the constant $\delta$ depends on the precise definition of the form factor and the values of the parameters $K$, $a$, $b$, $c$ and $n$ are summarised in Table\,\ref{tab:parameters} of Appendix\,\ref{sec:outer}. When the susceptibility $\chi_n(q_0^2)$ is evaluated at a non-vanishing value of $q_0^2$, the function $\phi$ must be multiplied by an additional factor\,\cite{Boyd:1997kz}, which, thanks to Eq.\,(\ref{eq:chi_n}), is given by $[t / (t - q_0^2)]^{(n+1)/2}$ on the unit circle. According to Eqs.\,(\ref{eq:substitutions}) its extension inside the unit disk leads to the final factor
\be
    \label{eq:q_0}
    \left( \frac{t}{z(t; 0)} \frac{z(t; q_0^2)}{t - q_0^2} \right)^{\frac{n+1}{2}} = \left( \frac{\sqrt{t_+ - t} + \sqrt{t_+}}{\sqrt{t_+ - t} + \sqrt{t_+ - q_0^2}} \right)^{n+1} ~ . ~
\ee

By means of the kinematical function\,(\ref{eq:outer}), the susceptibility $\chi_n(q_0^2)$ can be used as the upper bound $\chi^U$ to the quantity $\chi[f]$. 
In what follows we refer to
\be
    \label{eq:dispersive_bound}
   \chi[f] \leq  \chi^U = \chi_n(q_0^2)
\ee
as the {\em dispersive bound} to the form factor $f(z)$.
A more stringent estimate can be obtained by subtracting from $\chi_n(q_0^2)$ the contributions of the one-particle bound-states corresponding to the poles appearing in the Blaschke products. In Ref.\,\cite{Bourrely:1980gp} it was shown that for $q_0^2 \leq 0$ perturbative QCD can be applied to the calculation of the susceptibilities $\chi_n$ relevant in the case of the weak $V - A$ hadronic current.
The first non-perturbative calculations of the susceptibilities $\chi_n(q_0^2 = 0)$ have been carried out using LQCD simulations in Ref.\,\cite{Martinelli:2021frl}, and more recently in Refs.\,\cite{Melis:2024wpb, Harrison:2024iad}, in the case of the $b \to c$ transition, in Ref.\,\cite{DiCarlo:2021dzg} for the $c \to s$ transition and in Ref.\,\cite{Martinelli:2022tte} in the case of the $b \to u$ and $c \to d$ weak decays.

The basic step to develop such non-perturbative computations is to consider the Euclidean current-current correlator $V(\tau)$ depending on the Euclidean time-distance $\tau$  (i.e.\,the insertion of two currents $J$ integrated over their spatial separation)
\be
     \label{eq:Vtau_def}
     V(\tau) = \int d\vec{x} ~ \langle J^\dagger(\vec{x}, \tau) J(\vec{0}, 0) \rangle ~ , ~
\ee
where $\langle ... \rangle$ means the average of the $T$-product of the two currents $J$ over gluon and fermionic fields.
Such a correlator is related to the polarization function $\Pi(Q^2)$ evaluated at an Euclidean squared four-momentum $Q^2 \equiv - q^2$ by\,\cite{Blum:2002ii, Bernecker:2011gh}
\bea
    \label{eq:Vtau}
    V(\tau) & = & \frac{1}{2\pi} \int_{-\infty}^\infty dQ \, e^{-i Q \tau} Q^2 \Pi(Q^2) = \frac{1}{\pi} \int_{t_+}^\infty dt \, t  \, \mbox{Im}\Pi(t) \frac{1}{2\pi} \int_{-\infty}^\infty dQ \, 
                          \frac{e^{-iQ \tau}}{t+Q^2} \nonumber \\[2mm]
                & _{\overrightarrow{\tau > 0, \, t = \omega^2}} & \frac{1}{\pi} \int_{\sqrt{t_+}}^\infty d\omega \, \omega^2 \, \mbox{Im}\Pi(\omega^2) e^{- \omega \, \tau} ~ . ~  
\eea
Thus, the susceptibility $\chi_n(Q_0^2 = -q_0^2)$ can be obtained by evaluating the following (generalized) moment of the correlator $V(\tau)$:
\be
     \label{eq:chi_n_Q0}
     \chi_n(Q_0^2) \equiv \frac{2}{(2n)!!} \int_0^\infty d\tau \, \tau^{2n} V(\tau) \frac{j_{n-1}(Q_0 \tau)}{(Q_0 \tau)^{n-1}} ~ , ~ 
\ee
where $j_n(x)$ is the usual spherical Bessel function.
Indeed, since
\be
    \int_0^\infty d\tau \, \tau^{2n} e^{- \omega \tau} \frac{j_{n-1}(Q_0 \tau)}{(Q_0 \tau)^{n-1}} = (2n)!! \, \frac{\omega}{(\omega^2 + Q_0^2)^{n+1}} ~ , ~ \nonumber 
\ee
one gets
\be
    \chi_n(Q_0^2) = \frac{2}{\pi} \int_{\sqrt{t_+}}^\infty d\omega \, \frac{\omega^3}{(\omega^2 + Q_0^2)^{n+1}} \, \mbox{Im}\Pi(\omega^2) =
                               \frac{1}{\pi}\int_{t_+}^\infty dt \, \frac{t \, \mbox{Im}\Pi(t)}{(t + Q_0^2)^{n+1}} ~ . \nonumber 
\ee
At $Q_0^2 = 0$, since $\lim_{x \to 0} j_n(x)/x^n = 1/ (2n+1)!!$, one obtains
\be
    \label{eq:chi_n_Q0=0}
     \chi_n(Q_0^2 = 0) \equiv \frac{2}{(2n)!} \int_0^\infty d\tau \, \tau^{2n} V(\tau) ~ . ~ 
\ee

In almost all the applications of the dispersive bound $\chi_n(Q_0^2)$ the auxiliary variable $Q_0^2$ is chosen to be $Q_0^2 = 0$, mainly because of the possibility to estimate $\chi_n(Q_0^2 = 0)$ through perturbative QCD. We simply want to mention here that the impact of the $Q_0^2$-dependence of the susceptibility $\chi_n(Q_0^2)$, evaluated nonperturbatively on the lattice, has been investigated for the first time in Ref.\,\cite{Simula:2023ujs} in the case of the electromagnetic pion form factor.

\subsection{Unitary BGL expansion}
\label{sec:unitarity_BGL}

Once the kinematical function $\phi(z)$ is given by Eq.\,(\ref{eq:outer}) and the susceptibility $\chi_n$ is used as the dispersive bound $\chi^U$, the BGL expansion\,(\ref{eq:BGL}) can be applied to the case in which a set of input data $\{ f_i \}$, where $f_i \equiv f[z_i = z(t_i; t_0)]$ with $i =1, 2, ... N$, is known and one wants to obtain a model-independent prediction of $f(z)$ consistent with unitarity at a generic value of $z$ inside the unit disk.
To this end the usual procedure is to truncate the BGL expansion at some order $M$, namely
\be
    \label{eq:BGL_truncated}
    f^{(M)}(z) = \frac{\sqrt{\chi^U}}{\phi(z) B(z)} \sum_{k=0}^M a_k^{(M)} z^k ~ 
\ee
with\footnote{The use of a finite order $M$ in the truncated expansion\,(\ref{eq:BGL_truncated}) arises the issue of truncation errors, which were firstly discussed in Refs.\,\cite{Boyd:1995sq, Boyd:1997kz}. There, an upper limit to the truncation error was obtained, assuming that the coefficients $a_k^{(M)}$ coincide with the first $(M+1)$ coefficients $a_k$ of the untruncated BGL expansion (see Ref.\,\cite{Simula:2025fft}).}
\be
    \label{eq:unitarity_truncated}
    \sum_{k=0}^M \left[ a_k^{(M)} \right]^2 \leq 1 ~ . ~
\ee

The coefficients $a_k^{(M)} $ are typically obtained from a $\chi^2$-minimization procedure applied to the set of input data $\{ f_i \}$.
The implementation of the unitarity constraint\,(\ref{eq:unitarity_truncated}) directly in the fitting procedure is straightforward, even if it is not common to find it in many applications of the BGL expansion occurring in the literature. Either one can impose a penalty condition to be added to the $\chi^2$-variable or one can use the hyperspherical transformation described in the Appendix B of Ref.\,\cite{Simula:2023ujs}.
A further approach is the Bayesian procedure introduced recently in Ref.\,\cite{Flynn:2023qmi}, that allows to include in the truncated BGL expansion\,(\ref{eq:BGL_truncated}) some of the higher-order terms with $M > N$ in a way consistent with the unitarity constraint\,(\ref{eq:unitarity_truncated}).

One of the main goals of this work is to show that, in reality, in order to be consistent with unitarity there is a further hidden constraint beyond the usual one given by Eq.\,(\ref{eq:unitarity}), or by Eq.\,(\ref{eq:unitarity_truncated}) in the truncated version, that should be taken into account. To this end, let us consider the set of functions represented by all the BGL expansions\,(\ref{eq:BGL}) consistent both with the unitarity constraint\,(\ref{eq:unitarity}) and with a given set of input data $\{ f_i \}$. For sake of simplicity, in what follows we assume that the the form factor $f(z)$ is known for a series of distinct real values of $t = t_i < t_+$ (with $i =1, 2, ..., N$), so that all the $z_i$ are real and different from each other.
The untruncated BGL expansion\,(\ref{eq:BGL}) of the form factor $f(z)$ can be rewritten exactly in an additive form,  as derived firstly in Refs.\,\cite{Caprini:1980un, Caprini:1982av, Abbas:2010jc}. Here below, we use directly the final result of the computation described in Appendix\,\ref{sec:additive}, namely
\be
     \label{eq:additive}
     f(z) =  \frac{\sqrt{\chi^U}}{\phi(z) B(z)} \sum_{k=0}^\infty a_k z^k \equiv \beta(z) + R(z) ~ , ~
\ee
where
\be
    \label{eq:beta}
    \beta(z) = \frac{1}{\phi(z) B(z) d(z)} \sum_{i = 1}^N \phi(z_i) B(z_i) d_i f_i \frac{1 - z_i^2}{z - z_i}
\ee
is an analytic function of the real type reproducing exactly all the input data $\{ f_i \}$ with no free-parameters (i.e.\,$\beta(z) \to f_i$ when $z \to z_i$), while the {\em remainder} function $R(z)$, which should vanish for $z \to z_i$, is given by
\be
     \label{eq:rest}
    R(z) =  \frac{\sqrt{\chi^U - \chi[\beta]}}{\phi(z) B(z) d(z)} \sum_{k = 0}^\infty c_k z^k
\ee
with the constraints
\bea
    \label{eq:filter}
    \chi[\beta] & \leq & \chi^U ~ , ~ \\[2mm]
     \label{eq:unitarity_R}
     \sum_{k = 0}^\infty c_k^2 & \leq & 1.
\eea
In Eqs.\,(\ref{eq:beta})-(\ref{eq:rest}) the function $d(z)$ and the coefficients $d_i$ are given by\footnote{In other words the inverse of the function $d(z)$ is the product of the Blaschke factors corresponding to the locations $z_i$ of the known input data.}
\bea
   \label{eq:dz}
    d(z) & = & \prod_{m = 1}^N \frac{1 - z z_m}{z - z_m}  ~ , ~ \\[2mm]
    \label{eq:di}
    d_i & = & \prod_{m \neq i = 1}^N \frac{1 - z_i z_m}{z_i - z_m}  ~ , ~  
\eea
while in Eqs.\,(\ref{eq:rest})-(\ref{eq:filter}) the quantity $\chi[\beta]$ reads as
\be
      \label{eq:chi_DM}
      \chi[\beta] =  \sum_{i, j = 1}^N \phi(z_i) B(z_i) d_i f_i \, \phi(z_j) B(z_j) d_j f_j \, \frac{(1 - z_i^2) (1 - z_j^2)}{1 - z_i z_j} ~
\ee
and depends on the given set of input data $\{ f_i \}$.

The interpretation of the inequalities given by Eqs.\,(\ref{eq:filter}-\ref{eq:unitarity_R}) is as follows. The new inequality\,(\ref{eq:filter}) is necessary to guarantee that the function\,(\ref{eq:beta}) is consistent with unitarity and the remainder function\,(\ref{eq:rest}) is analytic of the real type.
Eq.\,(\ref{eq:filter}) represents a filter acting directly on the input data and allows to select only the subset $\{ \overline{f}_i \}$ of input data, which can be reproduced by BGL expansions satisfying the unitarity constraint\,(\ref{eq:unitarity}). Furthermore, the inequality\,(\ref{eq:unitarity_R}), acting on the coefficients $c_k$ of the remainder function $R(z)$, is a consequence of both the unitarity constraint $\chi[f] \leq \chi^U$, which is equivalent to the inequality\,(\ref{eq:unitarity}) on the coefficients $a_k$ of the BGL expansion, and the new inequality\,(\ref{eq:filter}).
Indeed, an explicit calculation, shown in Appendix\,\ref{sec:additive}, yields
\be
    \label{eq:chi_f}
    \chi[f] = \chi[\beta + R] = \chi[\beta] + \left( \chi^U - \chi[\beta] \right) \sum_{k = 0}^\infty c_k^2 ~ . ~
\ee
In other words the quantity $\chi[\beta]$ represents the minimum of $\chi[f]$ with $f$ given by Eq.\,(\ref{eq:additive}), so that, when Eq.\,(\ref{eq:unitarity_R}) is fulfilled, one has
\be
    \chi[\beta] \leq \chi[f] \leq \chi^U ~ . ~
\ee 

We stress that the above result implies two different unitarity constraints. The first one is given by $\chi[f] \leq \chi^U$, which implies the unitarity constraint\,(\ref{eq:unitarity}), which guarantees that all the BGL expansions\,(\ref{eq:additive}) are unitary. The other constraint is given by $\chi[\beta] \leq \chi^U$, which guarantees that the function\,(\ref{eq:beta}) fulfills unitarity. It tells us whether a given set of input data $\{ f_i \}$ is consistent with unitarity, i.e.\,whether it may be reproduced exactly by a BGL expansion\,(\ref{eq:BGL}) with coefficients $a_k$ satisfying the unitarity constraint\,(\ref{eq:unitarity}). When Eq.\,(\ref{eq:filter}) is not fulfilled, the input data $\{ f_i \}$ are not consistent with unitarity and should be removed from any analysis since they contain non-unitary effects\footnote{A consequence of the presence of non-unitary input data in a fitting procedure based on the BGL $z$-expansion is the value of the $\chi^2$-variable. Indeed, the latter one receives in general two contributions: the first one is related to the truncation order $M$ in Eq.\,(\ref{eq:BGL_truncated}), while the second one is due to the fact that non-unitary input data cannot be reproduced by a unitary BGL expansion. The latter contribution cannot be eliminated by increasing the order $M$ of the truncation and it is typically associated to a saturation of the dispersive bound $\chi^U$, i.e.\,$\chi[f] \simeq \chi^U$.}.
This may happen in various circumstances and we will describe some of them in a while. 
In all cases, starting from the given original sampling distribution for the input data, the application of the filter\,(\ref{eq:filter}) allows to construct the subset of input data consistent with unitarity. In this respect, we mention the importance sampling procedure, developed in Ref.\,\cite{Simula:2023ujs}, which must be adopted in the case of a large number of input data points for a proper application of the unitarity filter $\chi[\beta] \leq \chi^U$.

We point out that, strictly speaking, the use of the unitarity filter\,(\ref{eq:filter}) should not be viewed as a direct improvement of the BGL $z$-expansion\,(\ref{eq:BGL}) {\it per se}, but rather it represents an improvement of the standard unitary analysis based on the BGL $z$-expansion and applied to the reproduction of a given set of input data. 

Thus, in order to summarize the results of this Section a fitting procedure based on a unitary BGL $z$-expansion of the form factor $f(z)$ should include the following steps.
\begin{itemize}
\item[1.] Check the sample of input data $\{ f_i \}$, generated according to a given original sampling distribution probability, against the unitarity filter $\chi[\beta] \leq \chi^U$.
\item[2.] Select only the subset of filtered input data $\{ \overline{f}_i \}$ satisfying the unitarity constraint $\chi[\beta] \leq \chi^U$.
\item[3.] Apply the truncated BGL expansion\,(\ref{eq:BGL_truncated}) only to the unitary subset $\{ \overline{f}_i \}$ and impose on the coefficients $a_k^{(M)}$ the unitarity constraint\,(\ref{eq:unitarity_truncated}).
\end{itemize}

It often happens that a covariance matrix for the input data $\{ f_i \}$ is provided or assumed for lacking of direct information. In these cases a sample of data can be generated according to a Gaussian multivariate distribution. 
We stress that our procedure of filtering the input data is not limited at all to the case of Gaussian statistics, but it can be applied equally well to any kind of initial sampling distribution probability.
Analogously, the distribution of the filtered data $\{ \overline{f}_i \}$ may contain deviations from a Gaussian behaviour, which can be properly taken into account by a direct use of the filtered subset of input data.
In many phenomenological studies we have not observed sizable deviations from a normal statistics for the filtered subset $\{ \overline{f}_i \}$ (see for instance Fig.\,6 of Ref.\,\cite{Simula:2023ujs}), while non-Gaussian effects may definitely occur for the (high-order) coefficients of the unitary BGL $z$-expansion.
In any case it is recommended to present final results with error bars (and correlation matrix) only after checking that the distribution of the corresponding samples is basically Gaussian.

It is worthwhile to discuss briefly how non-unitary effects may plague the original input data $\{ f_i \}$. We address explicitly few selected circumstances.
\begin{itemize}
\item An interesting case is represented by the lattice calculations of the form factors relevant for the semileptonic $B \to D^* \ell \nu_\ell$ decays, performed in Refs.\,\cite{FermilabLattice:2021cdg, Harrison:2023dzh, Aoki:2023qpa}. The synthetic data points provided by each of the various lattice collaborations are the results of simulations carried out for several gauge ensembles at fixed values of the lattice spacing and for heavy-quark masses below the physical $b$-quark mass. Then, the extrapolations to the continuum limit and to the physical $b$-quark mass were performed assuming a simple polynomial dependence of the (HQET-inspired) form factors upon the recoil variable. Therefore, there is no guarantee that the momentum dependences of the form factors with definite spin-parity satisfy the corresponding unitarity constraints. In Refs.\,\cite{Martinelli:2021myh, Martinelli:2023fwm} we have applied to the lattice data of each collaboration the generalization of the unitarity filter\,(\ref{eq:filter}) to the case of the $B \to D^*$ transition and we have found that only a limited subset of the input data satisfies unitarity.
\item Another important case is represented by a dataset coming from an experiment, like in the case of the electromagnetic form factor of the pion analyzed in Ref.\,\cite{Simula:2023ujs}. Generally speaking, there is no reason why the systematic uncertainties should fulfill unitarity. Moreover, it may happen that statistical correlations among different data points are not addressed explicitly. The uncertainties provided for an experimental dataset may be viewed as a conservative estimate of the range of values of the form factor and, therefore, it should not be surprising if the application of the unitarity filtering selects a subset of the input data. Note that this does not necessarily imply the presence of effects due to operators beyond the Standard Model (SM). If the latter ones were instead present, part of them might lead to apparent violations of unitarity of the SM form factor. The investigation of such effects is well beyond the aim of the present paper and requires analyses at the level of cross sections or decay rates, where the $z$-expansion is not directly applicable.
\item In the case of multiple (experimental or theoretical) datasets, which are mutually uncorrelated, one can perform separate analyses for the individual input datasets and, after that, one can average in some way the individual results to compute a final prediction for the form factor (or for any observable of interest). Another option is to combine the different datasets in a unique larger dataset and check whether the final result for the form factor is consistent with the previous option. It is clear that independent fluctuations between different datasets may lead to violations of the unitarity constraint\,(\ref{eq:unitarity_truncated}) for the coefficients of the BGL $z$-expansion. The way to properly avoid that is to apply our filtering procedure to the full dataset (see for instance Ref.\,\cite{Martinelli:2023fwm}). 
\end{itemize}


Since the application of the unitarity filter\,(\ref{eq:filter}) can modify the sampling distribution probability for the input data, it is important to quantify and control its impact. To this end we introduce two quantities, $\Delta$ and $\epsilon$, defined as 
\be
      \label{eq:Delta}
      \Delta \equiv \left\{ \frac{1}{N} \sum_{i, j = 1}^N (\overline{f}_i - f_i) C_{ij}^{-1} 
                           (\overline{f}_j - f_j) \right\}^{1/2} ~_{\overrightarrow{C_{ij} = \sigma_i^2 \delta_{ij}}} ~ 
                           \left\{ \frac{1}{N} \sum_{i=1}^N \frac{(\overline{f}_i - f_i)^2}{\sigma_i^2} \right\}^{1/2} ~ , ~  
\ee
and
\be
      \label{eq:epsilon}
      \epsilon \equiv \left\{ \frac{1}{N} \sum_{i = 1}^N \frac{\overline{\sigma}_i^2}{\sigma_i^2} \right\}^{1/2} ~ ,  ~                   
\ee
where $\{ \overline{\sigma}_i \}$ represent the uncertainties of the ``filtered" dataset, i.e.\,the ones obtained after the application of the unitarity filter(s) to the input dataset.

The quantity $\Delta$ represents the average deviation of the new values $\{ \overline{f}_i \}$ from the starting ones $\{ f_i \}$, measured with respect to the starting covariance matrix $C_{ij}$. In other words, $\Delta < 1$ means that (on average) $\{ \overline{f}_i \}$ deviates from $\{ f_i \}$ by less than one standard deviation. The quantity $\epsilon$ is an estimate of how much the new uncertainties $\{ \overline{\sigma}_i \}$ deviate (on average) from the original ones $\{ \sigma_i  \}$. It can be smaller or larger than unity depending on whether $\{ \overline{\sigma}_i \}$ is (on average) smaller or larger than $\{ \sigma_i \}$.  If the application of the unitarity filter has no effect on the input dataset, one gets $\Delta = 0$ and $\epsilon = 1$. 

Thus, in order to avoid an anomalous impact of the unitarity filter(s) on both the mean values and errors of the input dataset we may put an upper limit $\Delta^*$ to $\Delta$ and a lower limit $\epsilon^*$ to $\epsilon$, namely:
\begin{itemize}
\item[a)] if $\Delta \leq \Delta^*$ and $\epsilon \geq \epsilon^*$, the``filtered" dataset can be kept and one can proceed with the computation of observables of interest for phenomenology;
\item[b)] if instead $\Delta > \Delta^*$ or $\epsilon < \epsilon^*$, one cannot proceed with a phenomenological study based on the filtered dataset.
\end{itemize}

\noindent Based on the values of $\Delta$ and $\epsilon$ found in the analyses of data on semileptonic decays of heavy mesons from Refs.\,\cite{DiCarlo:2021dzg, Martinelli:2021onb, Martinelli:2021myh, Martinelli:2022xir, Martinelli:2022tte, Martinelli:2023fwm} and on the pion electromagnetic form factor\,\cite{Simula:2023ujs}, we argue that acceptable values of $\Delta^*$ and $\epsilon^*$ are $\Delta^* \simeq 1$ and $\epsilon^* \simeq 0.5$.
In this way, if the condition $a)$ is satisfied, situations in which the average deviation of the new mean values of the FF exceed $1\sigma$ and the uncertainties are reduced by a factor of two or more are excluded. On the contrary, if the condition $b)$ is satisfied, one should avoid the use of the ``filtered" dataset for phenomenological applications. 
We expect the results obtained implementing our filtering procedure to be consistent with those corresponding to other methodologies of imposing unitarity, like those already proposed and discussed in Refs.\,\cite{Caprini:1980un, Caprini:1982av, Abbas:2010jc}.

Before closing this Section, we note that, since
\be
     \left| \sum_{k = 0}^\infty c_k z^k \right|^2 \leq \sum_{k = 0}^\infty c_k^2 \cdot \sum_{k^\prime = 0}^\infty |z|^{2 k^\prime} \leq \frac{1}{1 - |z|^2} ~ , ~
\ee
the additive form\,(\ref{eq:additive}) for real values of $z$ can be written as
\be
    \label{eq:fz_DM}
    \beta(z) - \sqrt{\gamma(z)} \leq f(z) \leq \beta(z) + \sqrt{\gamma(z)} ~ 
\ee 
with
\be
      \label{eq:gamma}
      \gamma(z) =  \frac{1}{1 - z^2} \frac{1}{\phi^2(z) B^2(z) \, d^2(z)} \left[ \chi^U - \chi[\beta] \right] ~ , ~
\ee
which is nothing else than the final result of the DM method of Ref.\,\cite{DiCarlo:2021dzg}, originally proposed in Refs.\,\cite{Bourrely:1980gp, Lellouch:1995yv}.
Without relying on any truncation or fitting procedure, Eq.\,(\ref{eq:fz_DM}) represents the uniform distribution of values of the form factor $f(z)$ at a generic value of $z$ consistent with unitarity as well as with the input data $\{ \overline{f}_i \}$. We stress that in the DM approach the unitarity filter $\chi[\beta] \leq \chi^U$ on the input data is always taken properly into account, since $\gamma(z)$ must be non-negative. The impact of the unitarity filter $\chi[\beta] \leq \chi^U$ has been addressed in the case of the semileptonic $B_{(s)} \to D_{(s)}^{(*)} \ell \nu_\ell$ decays in Refs.\cite{DiCarlo:2021dzg, Martinelli:2021onb, Martinelli:2021myh, Martinelli:2022xir, Martinelli:2023fwm},  for the semileptonic $B_{(s)} \to \pi(K) \ell \nu_\ell$ decays in Ref.\,\cite{Martinelli:2022tte} and for the pion electromagnetic form factor in Ref.\,\cite{Simula:2023ujs}. We highlight that the inclusion of the unitarity filter $\chi[\beta] \leq \chi^U$ may lead to important reduction of the uncertainties associated to the extrapolation of the form factors, especially in presence of a large number of input data\footnote{The changes in the mean values and errors on the FFs are always taken under control by evaluating the quantities $\Delta$ and $\epsilon$ given by Eqs.\,(\ref{eq:Delta}) and (\ref{eq:epsilon}), respectively.}. This is shown for instance in Fig.\,6 of Ref.\,\cite{Martinelli:2023fwm} in the case of the semileptonifc $B \to D^* \ell \nu_\ell$ form factors. There, the BGL results (green bands) are obtained without applying the unitary filtering\,(\ref{eq:filter}) to the input data and, therefore, they differ from the DM predictions (the red bands)\footnote{When the unitary BGL $z$-expansion is applied to the sample of {\it unitarity-filtered} input data, then the green bands coincide with the red ones, i.e.\,with the predictions of the DM approach.}.
We point out that an interesting generalization of Eq.\,(\ref{eq:chi_DM}), which includes also the first $K$ derivatives of the form factor at the origin $z = 0$, was derived in Ref.\,\cite{Caprini:1980un} (see also Ref.\,\cite{Abbas:2010jc}).

Finally, we comment briefly on the role of the auxiliary variable $t_0$, which fixes the value of $t$ at which $z(t_0; t_0) = 0$.
The value of $t_0$ may be chosen according to the requirement of accelerating as much as possible the convergence of the truncated BGL expansion\,(\ref{eq:BGL_truncated}). In Refs.\,\cite{Boyd:1994tt, Boyd:1995sq, Bourrely:2008za} it is argued that for semileptonic process $H_1 \to H_2 \ell \nu_\ell$ an optimal choice could be $t_0 = t_{opt} = (m_1 + m_2) \left( \sqrt{m_1} - \sqrt{m_2} \right)^2$, which minimizes the truncation error of the expansion\,(\ref{eq:BGL_truncated}) as estimated in Refs.\,\cite{Boyd:1995sq, Boyd:1997kz}. 
However, the form factor cannot depend on the choice of $t_0$. Since the Blaschke product $B(z)$ as well as the bound $\chi^U$ are independent of $t_0$ by construction, both the kinematical function $\phi(z)$ and the BGL coefficients $a_k$ must depend on $t_0$ in such a way to cancel any possible dependence of $f(z)$ on $t_0$. This implies that any difference in the truncated BGL $z$-expansions\,(\ref{eq:BGL_truncated}) corresponding to different values of $t_0$ may signal that the truncation error is not under full control.

\section{Multiple dispersive bounds}
\label{sec:seminclusive}

The quantity $\chi[f]$ is by definition an inclusive (positive) quantity, since all intermediate on-shell states $\overline{H}_1 H_2$ above the pair-production threshold $t_+ = (m_1 + m_2)^2$ are summed up. 
The second ingredient, proposed in this work, is the introduction of a set of limited, real and non-negative kernels $\widetilde{K}_p(z)$ (where $p=1,...,P$), constrained by 
\be
    \label{eq:sum_kernels}
    \sum_{p=1}^P \widetilde{K}_p(z) = 1 ~ . ~
\ee
In this way, the quantity $\chi[f]$ can be separated into a number of positive contributions $\chi_p[f]$, defined as 
\be
    \label{eq:chi_p}
    \chi_p[f] \equiv \frac{1}{2\pi i} \oint_{|z|=1} \frac{dz}{z} |\phi(z)f(z)|^2 \widetilde{K}_p(z) = \frac{1}{2\pi} \int_{-\pi}^\pi d\theta  |\phi(e^{i\theta})f(e^{i\theta})|^2 \widetilde{K}_p(e^{i\theta}) ~
\ee
with
\be
    \sum_{p = 1}^P \chi_p[f] = \chi[f] ~ . ~
\ee
The choice of the kernels $\widetilde{K}_p(z)$ is quite generic, provided they are limited, real and non-negative functions for $|z|=1$, i.e.\,along the branch-cut $t \geq t_+$. A possible choice is the squared modulus of an outer function, but, strictly speaking, there is no need to extend them inside the unit disk. They may even vanish on finite arcs of the unit circle. Indeed, a possible choice is represented by step functions into which the unit circle can be divided. Each step function corresponds to a well-defined range of masses for the intermediate $\overline{H}_1 H_2$ states, which may receive contributions from different resonances above the threshold $t_+$.

The quantities $\{ \chi_p[f] \}$ may contain important pieces of information on the behaviour of the absolute value of the form factor $f(z)$ along the branch-cut, which are usually {\em averaged} in the single, global quantity $\chi[f]$. In rough words, the same value of $\chi[f]$ may correspond to quite different behaviors of the form factor $f(z)$ along the branch-cut. Because of the dispersion relation\,(\ref{eq:DR}) this implies a spread of values of the form factor $f(z)$ at a generic value of $z$ consistent with the global unitary bound $\chi^U$. Instead, by indicating with $\chi_p^U$ an upper limit for the quantity $\chi_p[f]$, we expect that the simultaneous implementation of all the inequalities
\be
    \label{eq:multiple_bounds}
    \chi_p[f] \leq \chi_p^U \qquad p = 1, 2, ... P
\ee
can be more constraining on the form factor $f(z)$ at a generic value of $z$ than the use of the single, global inequality $\chi [f] \leq \chi^U$. The latter one can be always fulfilled, provided that 
\be
    \label{eq:sum_bounds}
    \sum_{p=1}^P \chi_p^U = \chi^U ~ . ~
\ee

An extensive numerical application of the idea of multiple dispersive bounds is presented in the companion paper\,\cite{Simula:2025fft}, where we show how the implementation of a double dispersive bound (i.e., with $P=2$) is crucial for an appropriate analysis of the effects of sub-threshold branch-cuts consistent with unitarity. In what follows, instead, we will focus on the formalism necessary to implement the multiple dispersive bounds in the BGL framework in the absence of sub-threshold branch-cuts.

\subsection{Implementation of the multiple dispersive bounds in the BGL expansion}
\label{sec:implementation}

Using the BGL expansion\,(\ref{eq:BGL}) in Eq.\,(\ref{eq:chi_p}) one gets
\be
     \chi_p[f] = \chi^U \, \sum_{k, k^\prime = 0}^\infty a_{k^\prime} U_{k^\prime k}^{(p)} a_k  \qquad p = 1, 2, ... P ~ , ~
\ee
where the matrix $U^{(p)}$ is given by
\be
     \label{eq:U_p}
     U_{k^\prime k}^{(p)} \equiv \frac{1}{2\pi} \int_{-\pi}^\pi d\theta \, e^{i (k - k^\prime) \theta} \widetilde{K}_p(e^{i \theta}) ~ . ~
\ee
The set of matrices $U^{(p)}$ are calculable once the set of kernels $\widetilde{K}_p(e^{i \theta})$ is specified, and satisfies the condition 
\be
    \label{eq:sum_Up}
    \sum_{p=1}^P U_{k^\prime k}^{(p)} = \delta_{k k^\prime} ~ . ~
\ee
The constraints $\chi_p[f] \leq \chi_p^U$ imply
\be
     \label{eq:multiple_BGL}
     \sum_{k, k^\prime = 0}^\infty a_{k^\prime} U_{k^\prime k}^{(p)} a_k \leq \frac{\chi_p^U}{\chi^U} \qquad p = 1, 2, ... P ~ . ~
\ee
Because of Eqs.\,(\ref{eq:sum_bounds}) and (\ref{eq:sum_Up}) the multiple constraints\,(\ref{eq:multiple_BGL}) are automatically consistent with the unitarity constraint\,(\ref{eq:unitarity}) on the coefficients $a_k$.
In practice, in the fitting procedure of the input data $\{ \overline{f}_i \}$ (i.e., the ones selected by the total unitarity filter $\chi[f] \leq \chi^U$) the constraints\,(\ref{eq:multiple_BGL}) can be applied simultaneously by adding appropriate penalty terms to the $\chi^2$-variable.

In the case of step functions, the unit circle can be divided in a set of arcs, namely
\be
     \widetilde{K}_p (e^{i \theta}) = \Theta[|\theta| - \alpha_{p-1}] -  \Theta[|\theta| - \alpha_p]
\ee
with $\alpha_0 = 0$ and $\alpha_P = \pi$. In this way, one gets
\be
     \pi U_{k^\prime k}^{(p)} = \frac{\mbox{sin}(k - k^\prime) \alpha_p}{k - k^\prime} - \frac{\mbox{sin}(k - k^\prime) \alpha_{p-1}}{k - k^\prime}  \qquad p = 1, 2, ... P ~ . ~
\ee

\subsection{Evaluation of the multiple dispersive bounds $\chi_p^U$}
\label{sec:chi_p}

At this point, it is necessary to discuss how the multiple bounds $\chi_p^U$ may be estimated in order to apply the constraints in Eq.\,(\ref{eq:multiple_BGL}). There are several strategies that may be pursued, as specified in what follows.
\begin{itemize}
\item[a)] One can develop models for the form factor $f(z)$ in the pair-production region (i.e.\,$t \geq t_+$).
\item[b)] The Operator Product Expansion (OPE) of the polarization function $\Pi(q^2)$ can be adopted for spacelike values of $q^2$.
\item[c)] Experimental data on the form factor $f(z)$ in the pair-production region can be used when available, like in the case of the electromagnetic pion form factor for timelike momenta from $e^+ e^-$ scattering data.
\item[d)] One can use the LQCD determination of the Euclidean current-current correlator $V(\tau)$.
\end{itemize}

We now briefly describe the last option, in which the kinematical function $\phi(z)$ is the outer function given by Eq.\,(\ref{eq:outer}). To this end, we introduce a set of limited, real and non-negative kernels $K_p(\tau)$ with $p = 1, 2, ... P$ given in terms of the Euclidean time-distance $\tau$. Again, the choice of such kernels is quite generic. In addition to the case of the step functions, an interesting choice is represented by {\em smoothed time windows}. A recent example is provided by the short-, intermediate- and long-distance windows introduced by the RBC Collaboration\,\cite{RBC:2018dos} for their studies of the Hadronic Vacuum Polarization (HVP) contribution to the muon $g - 2$. Recalling Eq.\,(\ref{eq:chi_n_Q0}), we use the kernel $K_p(\tau)$ to define the following dispersive bounds 
\be
     \label{eq:chi_p_U}
     \chi_p^U(Q_0^2) \equiv \frac{2}{(2n)!!} \int_0^\infty d\tau \, \tau^{2n} V(\tau)K_p(\tau) \frac{j_{n-1}(Q_0 \tau)}{(Q_0 \tau)^{n-1}} ~ . ~ 
\ee
Using Eq.\,(\ref{eq:Vtau}) one obtains
\be
     \label{eq:chi_p_bound}
     \chi_p^U(Q_0^2) = \frac{2}{\pi} \int_{\sqrt{t_+}}^\infty d\omega \, \frac{\omega^3}{(\omega^2 + Q_0^2)^{n+1}} \, \mbox{Im}\Pi(\omega^2) \widetilde{K}_p(\omega, Q_0^2) ~ , ~
\ee
where
\be
    \label{eq:kernel_LQCD}
    \widetilde{K}_p(\omega, Q_0^2) =  \frac{1}{(2n)!!} \, \frac{(\omega^2 + Q_0^2)^{n+1}}{\omega} \int_0^\infty d\tau \, \tau^{2n} e^{- \omega \tau} K_p(\tau) 
                                                            \frac{j_{n-1}(Q_0 \tau)}{(Q_0 \tau)^{n-1}} ~ . ~
\ee
Since by construction $\sum_{p = 1}^P K_p(\tau) = 1$, one has also $\sum_{p = 1}^P \widetilde{K}_p(\omega, Q_0^2) = 1$.

It is straightforward to realize that Eq.\,(\ref{eq:chi_p_U}) represents a unitary bound evaluated {\it nonperturbatively} on the lattice to the quantity $\chi_p(Q_0^2)$, defined by Eq.\,(\ref{eq:chi_p}), with $\widetilde{K}_p(e^{i\theta})$ corresponding to Eq.\,(\ref{eq:kernel_LQCD}) and
\be
    \label{eq:omega}
    \omega = \sqrt{t_+ + (t_+ - t_0) \, \mbox{cotg}^2(\theta / 2)} ~ . ~
\ee
At $Q_0^2 = 0$ one gets
\be
     \label{eq:chi_p_U_Q0=0}
     \chi_p^U(Q_0^2 = 0) \equiv \frac{2}{(2n)!} \int_0^\infty d\tau \, \tau^{2n} V(\tau)K_p(\tau) ~
\ee
and
\be
    \label{eq:kernel_LQCD_Q0=0}
    \widetilde{K}_p(e^{i \theta}) \to  \widetilde{K}_p(\omega, Q_0^2 = 0) = \frac{1}{(2n)!} \, \omega^{2n+1} \int_0^\infty d\tau \, \tau^{2n} e^{- \omega \tau} K_p(\tau) ~ . ~
\ee

\begin{figure}[htb!]
\begin{center}
\includegraphics[scale=0.525]{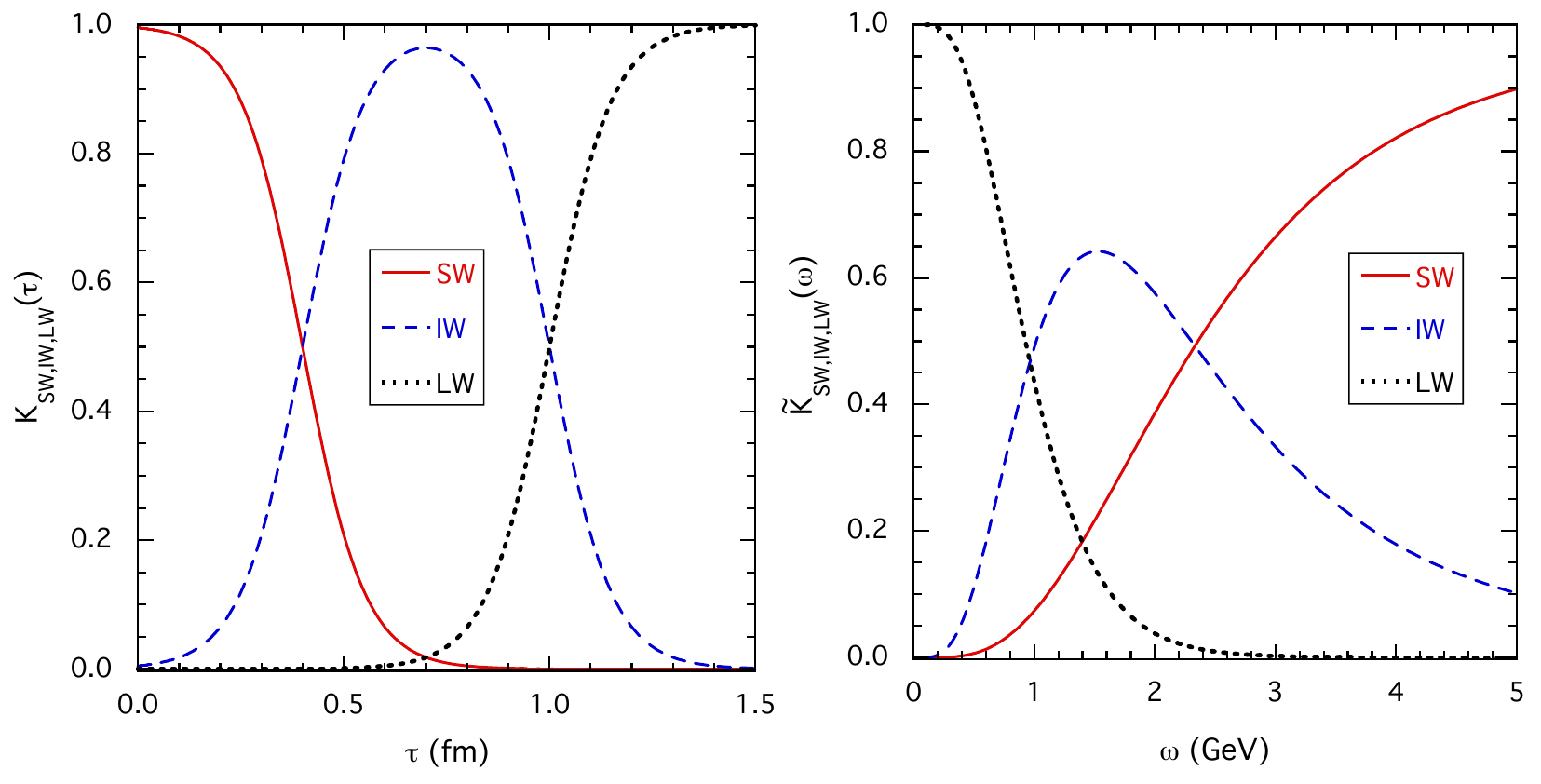}
\end{center}
\vspace{-0.75cm}
\caption{\it \small Left panel: the three Euclidean-time kernels $K_{SW}(\tau)$, $K_{IW}(\tau)$ and $K_{LW}(\tau)$ versus the time-distance $\tau$, given by Eqs.\,(\ref{eq:kernels_tau_RBC})-(\ref{eq:parms_RBC}) according to Ref.\,\cite{RBC:2018dos}. Right panel: the three corresponding energy kernels $\widetilde{K}_{SW}(\omega)$, $\widetilde{K}_{IW}(\omega)$ and $\widetilde{K}_{LW}(\omega)$ with $\omega$ given by Eq.\,(\ref{eq:omega}), obtained using Eq.\,(\ref{eq:kernel_LQCD_Q0=0}) at $Q_0^2 = 0$ and $n = 2$.}
\label{fig:RBC}
\end{figure}

For illustrative purposes only let us consider the short-, intermediate- and long-distance windows ($SW$, $IW$ and $LW$) introduced by the RBC Collaboration in Ref.\,\cite{RBC:2018dos}, namely
\bea
     \label{eq:kernels_tau_RBC}
     K_1(\tau) & = & K_{SW}(\tau) = 1 - \Theta(\tau, \tau_0; \Delta) ~ , ~ \nonumber \\[2mm]
     K_2(\tau) & = & K_{IW}(\tau) = \Theta(\tau, \tau_0; \Delta) -  \Theta(\tau, \tau_1; \Delta)~ , ~ \\[2mm]
     K_3(\tau) & = & K_{LW}(\tau) = \Theta(\tau, \tau_1; \Delta) ~ , ~ \nonumber
\eea
where the function $\Theta(\tau, \tau^\prime; \Delta)$ is defined as
\be
     \label{eq:theta_RBC}
     \Theta(\tau, \tau^\prime; \Delta) = \frac{1}{1+ e^{- 2(\tau - \tau^\prime) / \Delta}} ~ 
\ee
and the parameters $\tau_0$, $\tau_1$ and $\Delta$ are given by
\be
    \label{eq:parms_RBC}
    \tau_0 = 0.4 ~ \mbox{fm} ~ , ~ \qquad  \tau_1 =1.0 ~ \mbox{fm} ~ , ~ \qquad  \Delta = 0.15 ~ \mbox{fm} ~ . ~ 
\ee
The three Euclidean-time kernels $K_{SW}(\tau)$, $K_{IW}(\tau)$ and $K_{LW}(\tau)$ are shown in the left panel of Fig.\,\ref{fig:RBC}, while the right panel contains the corresponding {\em energy} kernels $\widetilde{K}_{SW, IW, LW}(\omega)$, evaluated using Eq.\,(\ref{eq:kernel_LQCD_Q0=0}) at $Q_0^2 = 0$ and $n = 2$.
Such energy kernels represent a possible choice for introducing multiple dispersive bounds in the analysis of the electromagnetic pion form factor $F_\pi^V(q^2)$ at spacelike values $q^2 \leq 0$. In this way, one can extend the analysis of the experimental data carried out  in Ref.\,\cite{Simula:2023ujs}, where only the \emph{total} dispersive bound $\chi^U(Q_0^2)$ was computed on the lattice. We leave the numerical study of such an extension to a future, separate work.

\section{Conclusions}
\label{sec:concl}

In this work we have proposed the introduction of two ingredients in the implementation of the widely used BGL $z$-expansion of hadronic form factors. 

On the one hand side, we have stressed the importance of the unitarity constraint given by Eq.\,(\ref{eq:filter}), which has a different interpretation w.r.t.\,the well-known unitarity constraint represented by Eq.\,(\ref{eq:upper}). In fact, Eq.\,(\ref{eq:filter}) corresponds to a filter dictated by unitarity and applied directly to a given set of input data. We have highlighted that the two constraints\,(\ref{eq:upper}) and (\ref{eq:filter}) allow to make the BGL $z$-expansion equivalent to the Dispersion Matrix method of Ref.\,\cite{DiCarlo:2021dzg}. 

On the other hand side, we have proposed the use of multiple dispersive bounds instead of a single, total dispersive bound through the introduction of a set of appropriate kernel functions in the evaluation of the hadronic susceptibilities. In this work we have focused on the implementation of multiple dispersive bounds in the BGL framework in the absence of sub-threshold branch-cuts. The numerical applications of multiple dispersive bounds depend strongly on the specific physical process of interest and, therefore, they will be left to future, dedicated works. The application of our proposal for exploiting multiple dispersive bounds to the case in which sub-threshold branch-cuts are present, is investigated in the companion paper\,\cite{Simula:2025fft}, where an extensive numerical application of double dispersive bounds is presented.

\section*{Acknowledgements}
We warmly thank Guido Martinelli for all the insightful discussions we had together and for his continuous support.
S.S.~is supported by the Italian Ministry of University and Research (MUR) under grant PRIN 2022N4W8WR.
L.V.~is supported by the Italian Ministry of University and Research (MUR) and by the European Union’s NextGenerationEU program under the Young Researchers 2024 SoE Action, research project ‘SHYNE’, ID: SOE\_20240000025.

\appendix

\section{Parameters of the outer functions for semileptonic meson decays $M_1 \to M_2 \ell \nu_\ell$}
\label{sec:outer}

We have collected in Table\,\ref{tab:parameters} the parameters $K$, $a$, $b$, $c$ and $n$ appearing in Eq.\,(\ref{eq:outer}) for the all form factors relevant for the semileptonic decays $M_1 \to M_2 \ell \nu_\ell$ of a pseudoscalar meson $M_1$ into a pseudoscalar or vector meson $M_2$ mediated by vector, axial-vector and tensor currents.
For the definition of the various form factors see, e.g., Ref.\,\cite{Bharucha:2010im}.

\begin{table}[htb!]
\renewcommand{\arraystretch}{1.2}
\begin{center}
\begin{tabular}{||c||c||c|c|c|c||c||}
\hline \hline
form factor & ~ current ~ & ~ $K$ ~ & ~ $a$ ~ & ~ $b$ ~ & ~ $c$ ~ & ~ $n$ ~ \\ \hline \hline
 $f_+$ & $V$ & 48 & 3 & 3 & 2 & 2 \\ \hline
 $f_0$ & $V$ & 16 & 1 & 1 & 1 & 1 \\ \hline
 $f_T$ & $T$ & 48 & 3 & 3 & 1 & 3 \\ \hline \hline
$g$ & $V$ & 96 & 3 & 3 & 1 & 2 \\ \hline
$f$ & $A$ & 24 & 1 & 1 & 1 & 2 \\ \hline
$\mathcal{F}_1$ & $A$ & 48 & 1 & 1 & 2 & 2 \\ \hline
$\mathcal{F}_2$ & $A$ & 64 & 3 & 3 & 1 & 1 \\ \hline
$T_0$ & $T$ & 48 & 1 & 1 & 1 & 3 \\ \hline
$T_1$ & $T$ & 24 & 3 & 3 & 2 & 3 \\ \hline
$T_2$ & $T$ & 24 & 1 & 1 & 2 & 3 \\ \hline \hline
\end{tabular}
\end{center}
\vspace{-0.25cm}
\caption{\it The parameters $K$, $a$, $b$, $c$ and $n$ appearing in Eq.\,(\ref{eq:outer}) for the all form factors relevant for the semileptonic decays $M_1 \to M_2 \ell \nu_\ell$ of a pseudoscalar meson $M_1$ into a pseudoscalar or vector meson $M_2$ mediated by vector (V), axial-vector (A) and tensor (T) currents.}
\label{tab:parameters}
\renewcommand{\arraystretch}{1.0}
\end{table}

\section{Additive form of the BGL expansion}
\label{sec:additive}

Inspired by the Blaschke factors\,(\ref{eq:Blaschke}), we introduce the following Blaschke-like product
\be
     \frac{1}{d(z)} \equiv \prod_{m = 1}^N \frac{z - z_m}{1 - z z_m}  ~, ~ \nonumber
\ee
which vanishes for $z \to z_m$, i.e.\, for all locations $z_m$ of the input data points. We remind that for sake of simplicity we assume that all $z_m$ are real with $|z_m| < 1$. The function $1 / d(z)$ is analytic of the real type inside the unit disk and unimodular on the unit circle.
When $z$ approaches $z_i$, one has
\be
      \prod_{m = 1}^N \frac{z - z_m}{1 - z z_m} = \frac{z - z_i}{1 - z_i^2} \frac{1}{d_i} + {\cal{O}}\left[ (z - z_i)^2 \right] ~ , ~ \nonumber
\ee
where
\be
     d_i = \prod_{m \neq i = 1}^N \frac{1 - z_i z_m}{z_i - z_m} ~ . ~ \nonumber
\ee
Thus, we can define the function $\psi_i(z)$ as
\be
    \psi_i(z) = \frac{1}{d(z)} \frac{d_i (1 - z_i^2)}{z - z_i} ~ , ~ \nonumber
\ee
which has the property $\psi_i(z_m) = \delta_{i m}$. Therefore, we can introduce the function
\bea
    \beta(z) & = & \frac{1}{\phi(z) B(z) d(z)} \sum_{i = 1}^N \phi(z_i) B(z_i) f_i \psi_i(z) \nonumber \\[2mm]
                 & = & \frac{1}{\phi(z) B(z) d(z)} \sum_{i = 1}^N \phi(z_i) B(z_i) d_i f_i \frac{1 - z_i^2}{z - z_i} ~ , ~ \nonumber
\eea
which is analytic of the real type inside the unit disk and reproduces exactly all the input data $\{ f_i \}$ with no free-parameters.

The first question is to establish whether the function $\beta(z)$ satisfies unitarity. To this end we have to calculate
\be
     \chi[\beta] = \frac{1}{2\pi i} \oint_{|z| = 1} \frac{dz}{z} \, \left| \phi(z) \beta(z) \right|^2 ~ . ~ \nonumber
\ee
Since both $B(z)$ and $d(z)$ are unimodular on the unit circle, one gets
\bea
    \chi[\beta] & = & \sum_{i, j = 1}^N  \phi(z_i) B(z_i) f_i \, \phi(z_j) B(z_j) f_j \, \frac{1}{2\pi i} \oint_{|z| = 1} \frac{dz}{z} \, \psi_i^*(z) \psi_j(z) \nonumber \\[2mm]
                     & = & \sum_{i, j = 1}^N  \phi(z_i) B(z_i) d_i f_i \, \phi(z_j) B(z_j) d_j f_j (1 - z_i^2) (1 - z_j)^2 \, \frac{1}{2\pi i} \oint_{|z| = 1} dz \, \frac{1}{(1 - z_i z) (z - z_j)} \nonumber \\[2mm]
                     & = & \sum_{i, j = 1}^N  \phi(z_i) B(z_i) d_i f_i \, \phi(z_j) B(z_j) d_j f_j \frac{(1 - z_i^2) (1 - z_j)^2}{1 - z_i z_j} ~ . ~ \nonumber
\eea
Thus, the function $\beta(z)$ is unitary only when $\chi[\beta] \leq \chi^U$, namely\footnote{An interesting generalization, which includes also the first $K$ derivatives of the form factor at the origin $z = 0$, was derived in Ref.\,\cite{Caprini:1980un} (see also Ref.\,\cite{Abbas:2010jc}).}
\be
     \sum_{i, j = 1}^N  \phi(z_i) B(z_i) d_i f_i \, \phi(z_j) B(z_j) d_j f_j \frac{(1 - z_i^2) (1 - z_j)^2}{1 - z_i z_j} \leq \chi^U ~ . ~ \nonumber
\ee
Let us now consider the remainder function $R(z)$, defined as
\be
    R(z) \equiv f(z) - \beta(z) ~ , ~ \nonumber
\ee
where $f(z)$ is given by the BGL expansion\,(\ref{eq:BGL}) with the coefficients $a_k$ satisfying the unitarity constraint\,(\ref{eq:unitarity}).
Since $\beta(z_i) = f_i$, the remainder function $R(z)$ vanishes for all $z = z_i$. Moreover, the product $\phi(z) B(z) R(z)$ must be analytic of the real type in the unit disk. Thus, we can write $R(z)$ in the form
\be
     R(z) = \frac{A}{\phi(z) B(z) d(z)} \sum_{k = 0}^\infty c_k z^k ~ , ~ \nonumber
\ee
where $A$ is a real constant. We now evaluate $\chi[f]$, obtaining
\bea
    \chi[f] = \chi[\beta + R] & = & \chi[\beta] + A^2 \sum_{k =0}^\infty c_k^2 \nonumber \\[2mm]
                                        & + & 2A \sum_{i =1}^N \phi(z_i) B(z_i) f_i \, \mbox{Re} \left\{ \frac{1}{2\pi i} \oint_{|z| = 1} \frac{dz}{z} \psi_i^*(z) \sum_{k = 0}^\infty c_k z^k \right \} ~ . ~ \nonumber
\eea
Thanks to the Cauchy theorem the last term in the r.h.s.\,of the above Equation vanishes, since
\be
     \frac{1}{2\pi i} \oint_{|z| = 1} dz \, \frac{1}{1 - z_i z} \, \sum_{k = 0}^\infty c_k z^k = 0 ~, ~  \nonumber
\ee
and, therefore, one gets $\chi[\beta + R] = \chi[\beta] + A^2 \sum_{k =0}^\infty c_k^2$. In order to satisfy unitarity one has $\chi[\beta + R] \leq \chi^U$, which implies
\bea
      A & = & \sqrt{\chi^U - \chi[\beta]} ~ , ~ \nonumber \\[2mm]
      \sum_{k = 0}^\infty c_k^2 & \leq & 1 ~ . ~ \nonumber
\eea

\bibliography{biblio}
\bibliographystyle{JHEP}

\end{document}